\documentclass[a4,aps,prb,superscriptaddress,twocolumn,floatfix,showpacs]{revtex4}
\usepackage{graphicx}
\usepackage[]{amsmath}
\usepackage{subfigure}

\begin{document}
\newcommand{\chem}[1]{\ensuremath{\mathrm{#1}}}

\title{Effect of magnesium doping on the orbital and magnetic order in \chem{LiNiO_2}}
\author{M.\ Bonda}
 \altaffiliation[New  address ]{\'{E}cole Polytechnique F\'{e}d\'{e}rale, Lausanne CH-1015,
Switzerland}

\author{M.\ Holzapfel}
\affiliation{ Grenoble High Magnetic Field Laboratory, CNRS,
BP166-38042 Grenoble cedex 9, France}

\author{S.\ de Brion}
 \email{sophie.debrion@grenoble.cnrs.fr}
\affiliation{ Grenoble High Magnetic Field Laboratory, CNRS,
BP166-38042 Grenoble cedex 9, France}
 \affiliation{Institut N\'{e}el / CNRS-UJF, BP166-38042
Grenoble cedex 9, France}

\author{C.\ Darie}
\affiliation{Institut N\'{e}el / CNRS-UJF, BP166-38042 Grenoble
cedex 9, France}

\author{T.\ Feher}
\affiliation{\'{E}cole Polytechnique F\'{e}d\'{e}rale, Lausanne CH-1015,
Switzerland}
\affiliation{Solids in Magnetic Fields Research Group
of the HAS and Institute of Physics, Budapest University of
Technology and Economics, P.O. Box 91, H-1521 Budapest, Hungary}

\author{P.\ J.\ Baker}
\affiliation{Clarendon Laboratory, University of Oxford,
Parks Road, Oxford OX1
3PU, United Kingdom}

\author{T.\ Lancaster}
\affiliation{Clarendon Laboratory, University of Oxford,
Parks Road, Oxford OX1
3PU, United Kingdom}

\author{S.\ J.\ Blundell}
\affiliation{Clarendon Laboratory, University of Oxford,
Parks Road, Oxford OX1
3PU, United Kingdom}

\author{F.\ L.\ Pratt}
\affiliation{ISIS Muon Facility, ISIS, Chilton, Oxon.
OX11 0QX, United Kingdom}

\date{\today}

\begin{abstract}
In \chem{LiNiO_2}, the Ni$^{3+}$ ions, with S=1/2 and twofold
orbital degeneracy, are arranged on a triangular lattice. Using
muon spin relaxation ($\mu$SR) and electron spin resonance (ESR),
we show that magnesium doping does not stabilize any magnetic or
orbital order, despite the absence of interplane Ni$^{2+}$. A
disordered, slowly fluctuating state develops below $12$~K. In
addition, we find that magnons are excited on the time scale of
the ESR experiment. At the same time, a $g$ factor anisotropy is
observed, in agreement with $\vert 3z^{2}-r^{2}\rangle$ orbital
occupancy.
\end{abstract}


 \maketitle

\section{\label{sec:Introduction}Introduction}

Orbital physics in oxides has attracted considerable interest
thanks to the discovery of high-temperature super-conductivity in
cuprates and colossal magnetoresistance in manganites
\cite{Tok00}. These macroscopic properties are based on strong
correlations between charge, orbital and magnetic degrees of
freedom. In this context, numerous studies have been performed on
the orbital and magnetic orders in the isostructural and
notionally isoelectronic compounds \chem{LiNiO_2}
\cite{GOO58,BRI05,CHU05}, \chem{NaNiO_2}
\cite{BON66,CHA00a,CHA00b,LEW05,BRI05,BRI07} and more recently
\chem{AgNiO_2} \cite{WAW07a,WAW07b,LAN07}. These compounds offer
the possibility of studying a triangular lattice of Ni$^{3+}$ ions
with spin $S=1/2$ and twofold orbital degeneracy ($e_g$ orbitals).
In \chem{NaNiO_2} and \chem{AgNiO_2}, it has been possible to
characterize both the orbital and magnetic ground states. However
in \chem{LiNiO_2}, these remain a matter of debate.

The octahedral oxygen crystal field at the \chem{Ni^{3+}} ions
lowers the original spherical SO$(3)$ symmetry of the Coulomb
field and  lifts partially the orbital degeneracy, leaving the
three fold $t_{2g}$ orbitals lower in energy than the $e_g$
orbital doublet. In \chem{NaNiO_2}, the Jahn-Teller (JT)-effect
further lifts the orbital degeneracy, giving rise to a ferro
orbital ordering of the $\vert 3z^{2}-r^{2}\rangle$ orbitals
\cite{CHA00b}. In \chem{AgNiO_2}, charge transfer occurs
3$e_g^{1}\rightarrow e_g^{2}+e_g^{0.5}+e_g^{0.5}$ with an
associated charge ordering \cite{WAW07a, WAW07b}. this process
removes the orbital degeneracy. In \chem{LiNiO_2}, there is no
experimental evidence for \emph{long-range} orbital ordering.
Evidence for a dynamic JT-effect \cite{BAR99} has been reported
while an EXAFS study \cite{ROU95} at room temperature concluded
rather that $\vert 3z^{2}-r^{2}\rangle$ orbitals are occupied.
More recently, this orbital occupancy  with no long range orbital
order has been confirmed by a neutron diffraction study
\cite{CHU05}.

 Magnetic ordering is affected by orbital ordering. In \chem{NaNiO_2},
a long range antiferromagnetic order is observed below $20$~K
\cite{BON66,CHA00b,DAR05,LEW05,BAK05} although the proposed A type
magnetic structure (ferromagnetic layers ordered
antiferromagnetically) cannot account for all the observed
magnetic excitations \cite{BRI07}. \chem{AgNiO_2} shows a more
complex antiferromagnetic order \cite{WAW07a,WAW07b,LAN07}, due to
different charges on the nickel ions: the \chem{Ni^{2+}} ($S=1$)
sublattice is arranged in ferromagnetic rows ordered
antiferromagnetically while the two other nickel sites carry a
small magnetic moment ($\leq 0.1 \mu_{\rm B}$). Surprisingly, no
long range magnetic order is present in \chem{LiNiO_2}. The
magnetic ground state has variously been described as a frustrated
antiferromagnet \cite{HIR91}, a spin glass \cite{YAM96} and a
disordered quantum state \cite{KIT98}.

It must be emphasized that, contrary to \chem{NaNiO_2} and
\chem{AgNiO_2}, \chem{LiNiO_2} is never stoichiometric and this
may affect the magnetic ground state: N\'u\~nez-Regueiro \emph{et
al.} \cite{NUN00} have shown that the presence of extra \chem{Ni^{2+}}
ions in the Li layers induces a ferromagnetic coupling between the
triangular planes which competes with the main antiferromagnetic
coupling leading to magnetic frustration. It may also affect the
orbital order since it generates also JT-inactive \chem{Ni^{2+}} ions
in the \chem{Ni^{3+}} layers\cite{PET06}. So how would a pure
\chem{LiNiO_2} sample behave? To address this issue, we have
investigated magnesium doped \chem{LiNiO_2},
 that is LiMg$_x$Ni$_{1-x}$O$_2$. Doping with magnesium is expected to
stabilize the lithium ions on their crystallographic sites and
thus to prevent insertion of nickel ions in lithium layers
\cite{POU00}. In this paper, we discuss  muon spin relaxation
($\mu$SR) and electron spin resonance (ESR) measurements performed
to investigate the magnetic and orbital order of magnesium-doped
\chem{LiNiO_2}. A comparison with quasi stoichiometric
\chem{LiNiO_2} is presented.

\section{Sample characterization}

Powder samples were synthetized  using the procedure described by
Pouillerie \emph{et al.} \cite{POU00} for
LiMg$_{x}$Ni$_{1-x}$O$_2$ and by Bianchi \emph{et al.}
\cite{BIA01} for \chem{LiNiO_2}. X-ray diffraction patterns were
collected using a Siemens 5000 diffractometer. Structural
refinement using the Rietveld method was performed with the
Fullprof program. All the samples crystallize in a rhombohedral
structure (space group $\textnormal{R}\bar{3}\textnormal{m}$) with
$\rm Ni^{3+}$ ions occupying the $3b\ (0\ 0\ 1/2)$ site and $\rm
Li^{+}$ ions the $3a\ (0\ 0\ 0)$ site. The crystal structure can
be imagined as an alternating stacking of lithium and nickel slabs
\cite{ROU95} with edge-sharing oxygen octahedra. Nickel ions are
arranged on a triangular lattice. Oxygen ions form an octahedron
around nickel sites which is responsible for lifting the orbital
degeneracy via the crystal field. The Rietveld refinements confirm
that the samples are deficient in lithium. In \chem{LiNiO_2},
\chem{Ni^{2+}} replaces \chem{Li^{+}}  \cite{BIA01}. In
LiMg$_{x}$Ni$_{1-x}$O$_2$, the detailed study by Pouillerie
\emph{et al} \cite{POU00} has shown that, for $x\geq0.2$, only
\chem{Mg^{2+}} replaces \chem{Li^{+}}. This is in agreement with
our X ray data as well as our magnetic susceptibility data (see
below). The resultant non stoichiometry in the Ni layers has been
described in ref.\cite{BIA01} and \cite{POU00}.

\begin{figure}[ht]
\includegraphics[width=8.5cm]{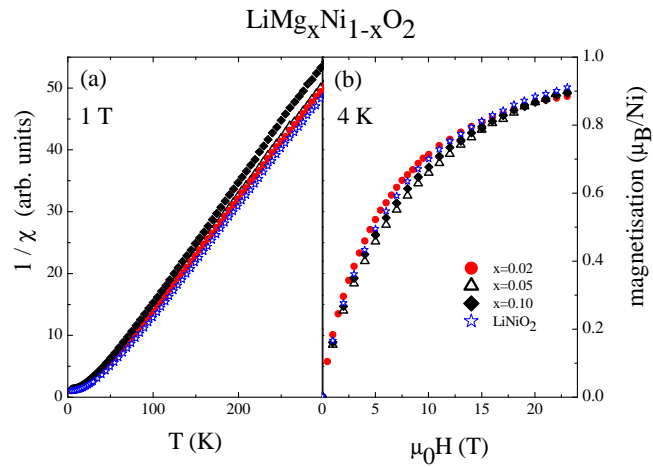}
\caption{(Color online) Magnetic properties normalized per Ni ion for
quasi-stoichiometric \chem{LiNiO_2} and LiMg$_x$Ni$_{1-x}$O$_2$ with
x=0.02, 0.05 and 0.10 . (a) Inverse of the susceptibility at 1T as
a function of temperature. (b) Magnetization at 4K as a function
of magnetic field.} \label{XT}
\end{figure}

\begin{table}[t]
\caption{Room temperature crystallographic parameters of the
$\textnormal{R}\bar{3}\textnormal{m}$ structure and magnetic
parameters of the Curie-Weiss law for quasi-stoichiometric and
Mg-doped \chem{LiNiO_2}.} \label{table1}

\begin{ruledtabular}
\begin{tabular}{ccccc}
 &\multicolumn{1}{c}{LiNiO$_{2}$}&\multicolumn{3}{c}{LiMg$_{x}$Ni$_{1-x}$O$_{2}$}\\
\hline & & {\it x}=0.02 & {\it x}=0.05 & {\it x}=0.10 \\
\hline
cell param.  & \textit{a}=2.8728 & \textit{a}=2.8724  & \textit{a}=2.8734 & \textit{a}=2.8773\\
(\AA) & \textit{c}=14.184 & \textit{c}=14.187 & \textit{c}=14.1978 & \textit{c}=14.2245\\
\hline
$\Theta_{\rm CW}$            & +29K     & +23.5K           & +22K   & +22K\\
\hline
$\Theta_{\rm CW}$ per \chem{Ni} &   +29K     &  +24K           &  +23K & +24.5K\\
\hline
$\mu_{\rm eff}$ per Ni     &   2.1 $\mu_{\rm B}$  &  2.1 $\mu_{\rm B}$ & 2.1 $\mu_{\rm B}$ &2.0 $\mu_{\rm B}$\\
\end{tabular}
\end{ruledtabular}
\end{table}

Structural and magnetic properties of the samples studied are
presented in Table \ref{table1}. Magnetic susceptibility
measurements $\chi(T)$ were  performed in a 1T magnetic field in
the temperature range 4K - 300K (Figure \ref{XT}(a)). All samples
present a Curie-Weiss behavior $\chi(T)=C/(T-\theta_{\rm CW})$ at
high temperature with $\theta_{\rm N}>0$ revealing dominant
ferromagnetic interactions (see table \ref{table1}). The effective
magnetic moment deduced from the slope $C$ remains close to
2.0$\mu_{\rm B}$ per Ni ion. The Curie-Weiss temperature
$\theta_{\rm CW}$ in Li$_{1-x}$Ni$_{1+x}$O$_2$ is a linear
function of the off-stoichiometry \cite{BIA01}: it extrapolates to
$\theta_{\rm CW}$=+24K for $x$=0 and can be used to determine $x$:
$x$=0.01 in our sample. For the magnesium doped samples, the
Curie-Weiss temperature is scarcely affected by the number of Mg
ions. It is also quite close to the value expected for pure
\chem{LiNiO_2} and reduced compared to our \chem{LiNiO_2}. This is
a confirmation that the additional ferromagnetic interaction
present in non-stoichiometric \chem{LiNiO_2} is absent in
magnesium doped \chem{LiNiO_2}. This makes these samples
particularly interesting to determine the orbital and magnetic
ground states of pure \chem{LiNiO_2}.

Magnetization measurements were performed at 4.2 K in magnetic
fields up to 23 T (Figure \ref{XT}(b)). Here also a common
behavior for all the samples is observed: a smooth increase of the
magnetization and absence of saturation as in a spin glass. This
is the same behavior as in quasi stoichiometric \chem{LiNiO_2}. In
the following, we will present $\mu$SR and ESR results obtained on
$5~\%$ Mg  doped sample with those on quasi stoichiometric
\chem{LiNiO_2}.

\section{\label{musrresults} $\mu$SR results}

Our muon-spin relaxation ($\mu$SR) experiments~\cite{blundell99}
were carried out using the GPS instrument at the Paul Scherrer
Institute (PSI), Villigen, Switzerland. Spin-polarized positive
muons ($\mu^{+}$, mean lifetime $2.2~\mu$s, momentum $29$~MeV$/c$,
gyromagnetic ratio $\gamma_{\mu}/2\pi = 135.5$~MHz~T$^{-1}$) were
implanted into the bulk of our polycrystalline samples. Dipole
field calculations carried out for the structurally similar
compound \chem{NaNiO_2} suggest that the muon stopping sites are
near the oxygen ions that form the octahedra around the \chem{Ni^{3+}}
ions~\cite{BAK05}. The muons stop quickly in the samples
(within $\sim 1$~ns) without significant loss of spin polarization,
and their average spin polarization is measured as a function of
time using the asymmetry, $A(t)$, of positrons emitted by muons
decaying within the sample.

\begin{figure}[ht]
\includegraphics[width=8.5cm]{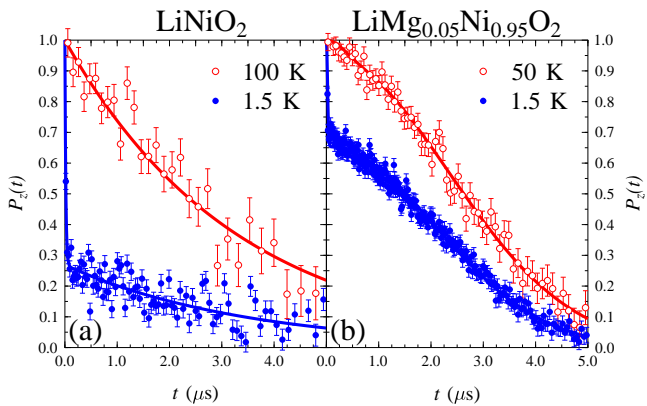}
\caption{(Color online)
Corrected asymmetry [$P_{z}(t)$] data at high and low temperature for:
(a) \chem{LiNiO_2} with fits to Eq.~\ref{purefit} using the parameters
shown in Fig.~\ref{musrPure}, and
(b) 5~\% Mg-doped sample with fits to Eq.~\ref{mgfit} using the
parameters shown in Fig.~\ref{musrMg}.
For clarity of display the histogram binning has been altered and
the data sets have been normalized with a temperature independent
background subtracted.
}
\label{musrraw}
\end{figure}

The measured positron asymmetry, $A(t)$, was corrected for the
non-relaxing background signal resulting from muons stopping in
the cryostat and sample holder and normalized to unity. The corrected
asymmetry, $P_{z}(t)$, is plotted in Fig.~\ref{musrraw} for both
compounds. It is clear that in both samples there is no spin precession
of the implanted muons, which would give rise to coherent oscillations
in the asymmetry spectra. Together with the fact that the high and low
temperature spectra relax to the same background asymmetry, this is
strong evidence for a lack of long range magnetic order in either
sample, as already suggested by the magnetization data (Fig.~\ref{XT}).
The form of $P_{z}(t)$ is sensitive to spin fluctuations on timescales
between approximately $10^{-12}$ and $10^{-4}$~s.

Examining the data plotted in Figure~\ref{musrraw}, we see that in
both compounds the low temperature data is composed of two
relaxing components, one with a much higher relaxation rate than
the other. The previous $\mu$SR study of \chem{LiNiO_2} used a
stretched exponential $P_{z}(t) \propto \exp[-(\lambda t)^k]$ to
describe the asymmetry data~\cite{JPCM.17.1341}, but this
parametrization was not able to describe our data for either
compound over the entire measured temperature range. Two relaxing
components were used by \textcite{BAK05} to describe the muon spin
relaxation at temperatures just above $T_{\rm N}$ in
\chem{NaNiO_2}, with the amplitude of the fast relaxing component
increasing as $T_{\rm N}$ was approached. Coupled with ac
susceptibility measurements, this was taken to be evidence for
coalescing magnetic clusters preceding the onset of long range
magnetic order.  In both of the present samples such long range
magnetic order never sets in. The form of the slower relaxing
component in each sample is distinguishable, as can be seen in
Figure~\ref{musrraw}, and is the same at high and low temperature.
We can also see that there is a significant difference in the
proportion of the fast and slow relaxing components of the
asymmetry between the two samples. This is due to a difference in
the quasistatic magnetic volume fraction (see below).We now go on
to discuss the results from each sample in turn.

\begin{figure}[t]
\includegraphics[width=8.5cm]{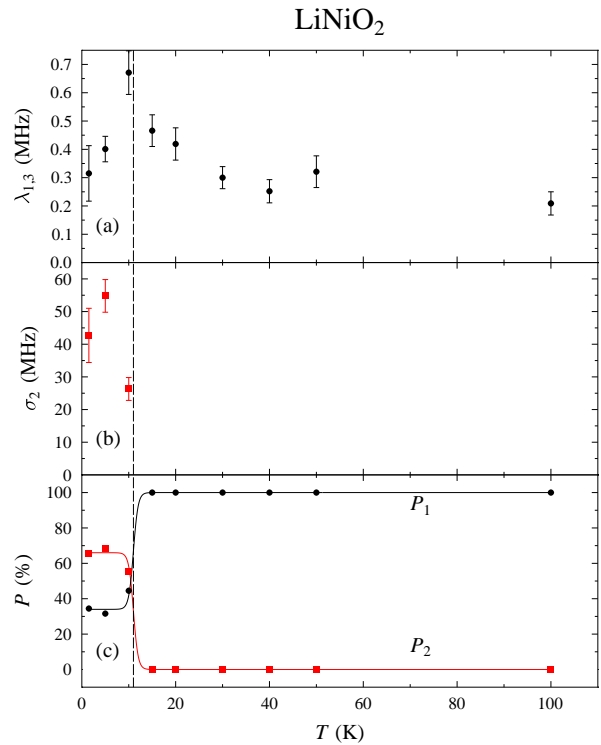}
\caption{(Color online) Parameters derived from fitting
Eq.~\ref{purefit} to the raw positron asymmetry data for
quasi-stoichiometric \chem{LiNiO_2}. (a) $\lambda_1$ and
$\lambda_3$ describing the slower relaxing component. (b)
$\sigma_2$ describing the faster relaxing component. (c)
Amplitudes of the relaxation components $P_1$ and $P_2$. The lines
are guides to the eye.} \label{musrPure}
\end{figure}

Below approximately $11$~K, the asymmetry signal in quasi-stoichiometric
\chem{LiNiO_2} is well described by the function:
\begin{equation}
P_{z}(t) = P_{1}e^{-\lambda_{1}t} + P_{2}e^{-(\sigma_{2}t)^{2}},
\label{purefit}
\end{equation}
where the exponential component describes the slow relaxing component of
the signal and the Gaussian component describes the fast relaxation seen
at short times (Figure~\ref{musrraw}). It is difficult to resolve the
exact form of the fast relaxing component, in either sample, because the
relaxation rate is so large.

At low temperature $P_1 \sim 1/3$ and $P_2 \sim 2/3$, which is the
form expected for the relaxation observed in a powder sample in
the slow fluctuation (quasistatic) limit. If the sample exhibited
long range magnetic order $P_2$ would be multiplied by an
oscillating function, as seen in \chem{NaNiO_2}~\cite{BAK05} or
\chem{AgNiO_2}~\cite{LAN07}. The ratio of $P_{1}$ and $P_2$ and
the lack of muon precession suggests that at low temperature we
have quasistatically disordered magnetic moments throughout the
sample. The disorder in the magnetic moments prevents the
observation of coherent muon precession. From the size of the
Gaussian relaxation rate ($\sigma = {\gamma_{\mu}\Delta
B}/\sqrt{2}$) we can estimate that the distribution of magnetic
fields is approximately $\Delta B \sim 0.5$~T.

It is possible to describe the asymmetry data successfully with
equation~\ref{purefit} up to around $11$~K, where we observe a
sharp crossover to a regime where the muon spin relaxation is well
described by a single exponential $e^{-\lambda_{3}t}$, typical of
paramagnetic spin fluctuations when the fluctuation rate is fast
compared with the width of the magnetic field distribution. The
crossover is accompanied by a peak in $\lambda$, seen in
Figure~\ref{musrPure}, associated with the slowing down of the
electronic fluctuations. \textcite{JPCM.17.1341} found a similar
peak in their measured relaxation rate in an applied longitudinal
field of $0.6$~T, as would be expected if the field has decoupled
the muon relaxation due to the quasistatic moments. Their zero
field results~\cite{JPCM.17.1341} show a sharp increase in the
relaxation rate of the stretched exponential used to parametrize
the data below $10$~K. The magnitude of the relaxation rate found
at low temperature is similar to the value of $\sigma_{2}$ we
find. Parametrizing our data in terms of two separate components
(Equation~\ref{purefit}) allows us to separate these contributions
to the muon relaxation in zero applied field. From the magnitude
of $\lambda_3$ at high temperature, we can estimate the electronic
fluctuation rate in the paramagnetic phase to be $\tau = \lambda_3
/ {2 \gamma^{2}_{\mu} \langle \Delta B^{2} \rangle} = 32$~ps.

\begin{figure}[t]
\includegraphics[width=8.5cm]{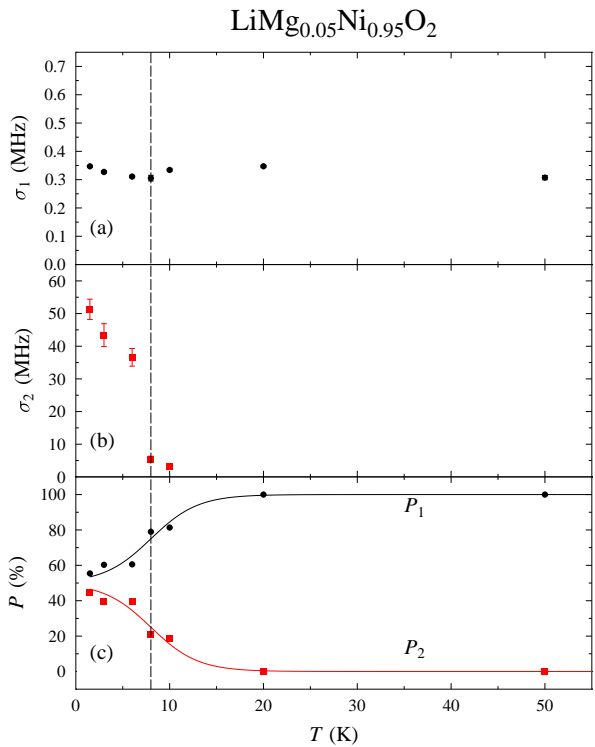}
\caption{(Color online) Parameters derived from fitting
Eq.~\ref{mgfit} to the raw positron asymmetry data for 5~\%
Mg-doped \chem{LiNiO_2}. (a) $\sigma_1$ describing the slower
relaxing component. (b) $\sigma_2$ describing the faster relaxing
component. (c) Amplitudes of the relaxation components $P_1$ and
$P_2$. The lines are guides to the eye. } \label{musrMg}
\end{figure}

In \chem{LiMg_{0.05}Ni_{0.95}O_2} the behavior observed is quite similar to the
quasi-stoichiometric \chem{LiNiO_2} case. Again we see two separate components
to the muon spin relaxation at low temperature, but in this sample both components
take a Gaussian form. This can be described as:
\begin{equation}
P_{z}(t) = P_{1}e^{-(\sigma_{1}t)^{2}} + P_{2}e^{-(\sigma_{2}t)^{2}},
\label{mgfit}
\end{equation}
where we take $\sigma_{1} < \sigma_{2}$. We must consider why the
slow relaxing component is Gaussian rather than exponential in
this case, as it is also at high temperature. This appears to be
due to the motional narrowing of the electronic fluctuations,
which are sufficiently fast to leave the muon time window, so that
the muon is sensitive to the Gaussian distribution of magnetic
fields due to randomly orientated nuclear dipoles. This is
consistent with the small and almost temperature independent slow
Gaussian relaxation $\sigma_{1}$, plotted in
Figure~\ref{musrMg}(a). Below $\sim 8$~K we do not observe a $2:1$
ratio in the amplitudes of the two components, so the slowly
fluctuating region of the sample is not occupying the full sample
volume. From the ratio $P_{2}:P_{1}$ we can estimate that slowly
fluctuating moments occupy approximately three-quarters of the
sample volume. The $\mu$SR measurements suggest that the
electronic moments in the rest of the sample fluctuate
sufficiently fast that the slow, temperature-independent
relaxation is due to the nuclear moments. The magnitude of
$\sigma_{2}$ is similar to that observed in the
quasi-stoichiometric sample, so the distribution of magnetic
fields at the muon stopping site is similar, $\Delta B \sim
0.5$~T.

From the $\mu$SR data we can conclude that Mg-doping has a
significant effect on the spin dynamics of \chem{LiNiO_2}. While
neither sample shows long-ranged magnetic order, as the
magnetization data confirms (Fig.~\ref{XT}), the low-temperature
state is affected by the presence of the \chem{Mg} ions. The
quasi-stoichiometric sample shows a sharp crossover to a
disordered and slowly fluctuating state throughout the sample
volume below $12$~K. This is accompanied by a peak in the
relaxation rate $\lambda_1$, associated with the slowing down of
electronic fluctuations. The 5~\% Mg-doped sample enters a similar
ground state at low temperature, but one that only occupies around
three-quarters of the sample volume, the remaining part fluctuates
too fast to be detectable, and only the nuclear origin of the muon
depolarization is observable. Despite the different
parametrization, the results of the previous $\mu$SR
study~\cite{JPCM.17.1341} of \chem{Li_{0.98}Ni_{1.02}O_{2}} appear
to be intermediate between those of our doped and undoped samples.
Their sample's almost Gaussian high-temperature relaxation is
quite similar to that in our $\%5$ Mg-doped sample, and the
magnitude of the low-temperature relaxation rate is similar to
that of the fast relaxing components observed in both of our
samples. The 0.6T longitudinal field measurements give a
relaxation rate with a temperature dependence quite similar to the
slow relaxing component in our \chem{LiNiO_2} sample. This is in
agreement with our assignment of the slow relaxing component to
fast fluctuating moments that would not be decoupled by such a
field. These comparisons suggest that Chatterji et al's sample is
likely to have had a slightly higher concentration of
substitutional defects than our \chem{LiNiO_2} sample, but lower
than our $5\%$ Mg-doped sample.

\section{ESR results}

\begin{figure}[ht]
\includegraphics[width=8.5cm]{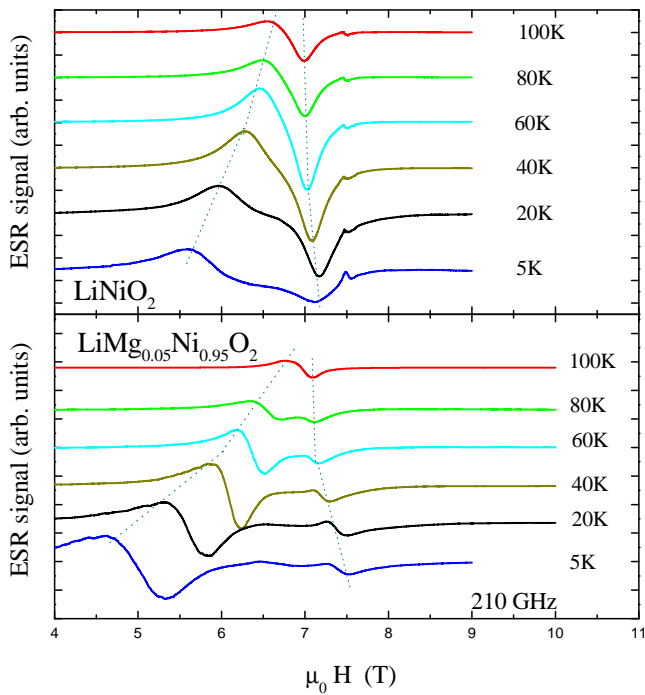}
\caption{(Color online) ESR spectra at 210 GHz for quasi
stoichiometric and Mg doped \chem{LiNiO_2}.} \label{fig:ESRspec}
\end{figure}

ESR measurements of magnetic \chem{Ni^{3+}} ions were carried out
over a temperature range of 4 to 200~K and at three different
frequencies, that is, 210, 314 and 420~GHz, using a quasi optical
bridge and a 14~T superconducting magnet. A field modulation was
used so that the derivative of the ESR absorption is recorded
(Fig.\ref{fig:ESRspec}).

\begin{figure}[ht]
\includegraphics[width=8.5cm]{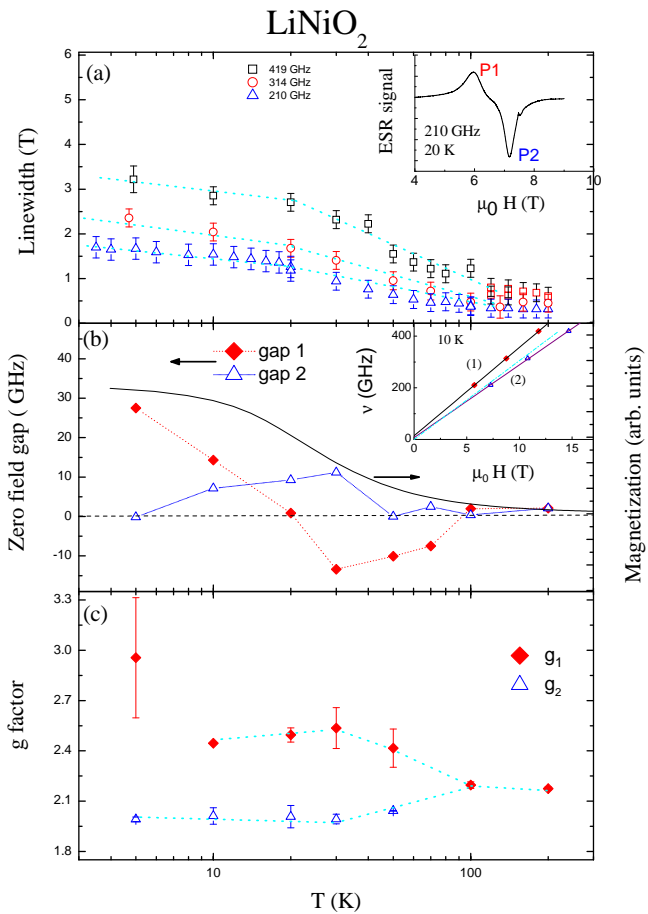}
\caption{(Color online) ESR spectra parameters for quasi
stoichiometric \chem{LiNiO_2} as a function of temperature.}
\label{fig:LiNiOesr}
\end{figure}

\begin{figure}[ht]
\includegraphics[width=8.5cm]{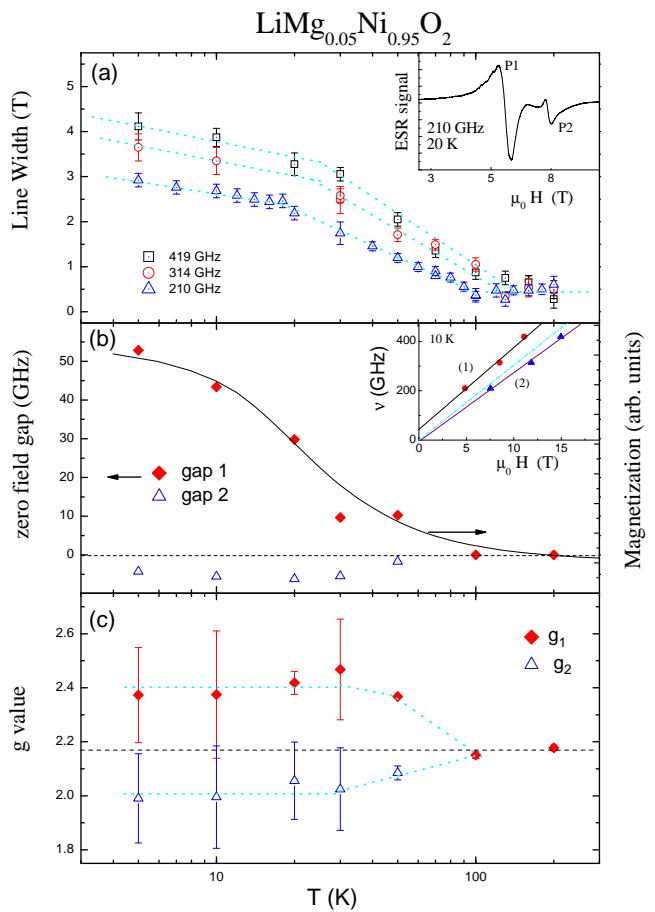}
\caption{(Color online) ESR spectra parameters for Mg doped
\chem{LiNiO_2} as a function of temperature.}
\label{fig:LiMgNiOesr}
\end{figure}

In highly correlated systems ESR absorption presents a large
linewidth; measuring at high frequencies provides better
resolution and also makes it possible to follow magnetic modes up
to high magnetic fields. Both the quasi-stoichiometric and
Mg-doped \chem{LiNiO_2} samples exhibit a similar ESR response.
Three different temperature regimes are observed (see Figs.
\ref{fig:ESRspec}-\ref{fig:LiMgNiOesr}). Above about 100~K, the
ESR absorption has a Lorentzian lineshape, in particular the
recorded derivative spectrum is symmetric and its linewidth
remains nearly constant for all three frequencies. Below
$\approx100$~K, the ESR signal becomes asymmetric and widens with
increasing frequency, reflecting some kind of anisotropy,
crystalline or magnetic in origin. This anisotropy increases
dramatically as the temperature is lowered. This is particularly
clear in Fig.~\ref{fig:LiNiOesr} and \ref{fig:LiMgNiOesr} where
the ESR spectrum total linewidth is plotted as a function of
temperature for the three different frequencies. A third
temperature regime occurs below about 30~K where the linewidth
levels off. Note that the ESR spectra are always wider in the Mg
doped sample than in the quasi stoichiometric one. Their
lineshapes at low temperature differ slightly. We will concentrate
on the extremal features (peaks P1 and P2) which are similar in
both powdered samples.

The question now is to determine the origin of the broadening of
the ESR signals with temperature. There are two possibilities.
Either it arises from spin-orbital coupling causing an anisotropy
of the $g$-factor when the expected JT effect occurs as observed
in \chem{NaNiO_2} \cite{CHA00b}, or it arises from magnetic
excitations when some magnetic order is present. Both processes
may occur simultaneously. To go further into the analysis, we use
a frequency ($\nu$)/magnetic field ($H$) diagram where the extrema
of the ESR spectra are plotted. This procedure enables us to
determine the main ESR absorption processes which occur in the
powdered samples. Using a linear approximation, we can calculate
the corresponding $g$-factors from the high field slope and the
zero field gaps due to magnetic excitations. This approximation is
valid at high fields (see for instance
Ref.~\onlinecite{BRI07}).The corresponding frequency/field diagram
and linear fit are shown at 5~K  on Fig.~\ref{fig:LiNiOesr} for
quasi stoichiometric \chem{LiNiO_2} and on
Fig.~\ref{fig:LiMgNiOesr} for Mg doped \chem{LiNiO_2}. The
light-blue line in the middle is the one calculated for the
paramagnetic $g$-factor of free electrons ($g_0=2.003$, no gap).
This linear procedure is then done for several temperatures. As
the quality of spectra may vary from one frequency to another
error bars may be wide. Above 100~K, the $g$-factor is unique with
the same value for both samples quite close to the one derived for
the purest \chem{LiNiO_2}: $g=2.17$ \cite{CHA02}. While spectra
are broadening below 100~K, two distinguishable branches are
observed with different $g$-factor and zero field gap. Below 30~K
both $g$-factors stay almost constant while the zero field gaps
continue to vary, at least for Mg doped \chem{LiNiO_2}. From this
analysis, it is clear that two processes occur simultaneously, in
both samples: the $g$ factor becomes anisotropic \textit{and}, at
the same time, magnetic excitations develop with a significant
zero field gap for one branch (30-50 GHz at 5K).

 If one assumes a crystallographic origin for the $g$ anisotropy, one can
calculate the  \emph{effective} $g$-factor of the spin hamiltonian
within perturbation theory applied to the symmetry considered
\cite{IBE62}. For an elongation of the oxygen octahedra (occupied
orbitals $\vert3z^{2}-r^{2}\rangle$) :
\begin{align}
g_\parallel&=g_0\\
g_\perp&=g_0-\frac{6\lambda}{\Delta_{cf}}\quad,
\end{align}
whereas for a flattening of the octahedra (occupied
orbitals $\vert3x^{2}-y^{2}\rangle$):
\begin{align}
g_\parallel&=g_0-\frac{8\lambda}{\Delta_{cf}}\\
g_\perp&=g_0-\frac{2\lambda}{\Delta_{cf}}\quad
\end{align}
where $\lambda$ is the spin-orbit coupling and $\Delta_{cf}$ the
crystal field splitting. Since $\lambda < 0 $ for a more than half
filled ion as Ni$^{3+}$, we expect $g_\parallel< g_\perp$ for
elongated octahedra and $g_\parallel > g_\perp$ for flattened
octahedra. In a powder spectrum, all the orientations are equally
probable giving rise to a wide spectrum with a particular
lineshape for each case. It was previously shown that in
\chem{NaNiO_2}, a powdered spectrum with $g_\parallel < g_\perp$
is observed at 200~K while a single value of $g$ is observed in
\chem{LiNiO_2} \cite{CHA02} . This is confirmed here for our
\chem{LiNiO_2} sample as well as the Mg doped sample above 100~K
with the same unique value $g=2.17$.


 Below $\sim 100K$, two different values of $g$ are
observed: $g_1=2.0$ and $g_2=2.4$. Since $g_1$ is equal to $g_0$,
as in the model with occupied $\vert3z^{2}-r^{2}\rangle$)
orbitals, our results are slightly in favor of this model. This is
the same orbital occupancy as in \chem{NaNiO_2} \cite{CHA00b}.
However, in the orbital sector, the Li compounds differs from that
in the Na compound in at least two ways. First, in \chem{NaNiO_2},
there is a ferro-orbital ordering (cooperative JT effect) below
480~K leading to a monoclinic unit cell. In Li compounds, no long
range orbital order is observed: the unit cell remains
orthorhombic. Second, for the local orbital occupancy, it is clear
experimentally in \chem{NaNiO_2} that the $\vert
z^{2}-r^{2}\rangle$ orbitals are occupied at low temperature
\cite{CHA00b}. In Li compounds, the situation is more
controversial. An ESR study \cite{BAR99} suggests  that a
dynamical Jahn-Teller effect exists that becomes static with
$g_\parallel > g_\perp$ at low temperature and high magnetic
field, which means $\vert 3z^{2}-r^{2} \rangle$  orbital
occupancy. An EXAFS study \cite{ROU95} at room temperature implies
the presence of elongated oxygen octahedra around the Ni ions, in
agreement with the $\vert 3z^{2}-r^{2} \rangle$ orbital occupancy.
More recently, a neutron diffraction and pair density function
analysis \cite{CHU05} concludes that local orbital order occurs
below 375~K with the $\vert 3z^{2}-r^{2} \rangle$ orbitals
organized in trimers with no long range order. How can we
reconcile all these different results? Note first that our ESR
spectra are similar to those observed by Barra et al: a single
line at high temperature and a splitted spectrum at low
temperature. Our more detailed frequency study shows that the
associated g-anisotropy corresponds rather to the $\vert
z^{2}-r^{2}\rangle$ orbital occupancy and that there is also an
opening of a zero field gap, which reveals the presence of
magnons. Then our results at low temperature are in agreement with
the EXAFS and neutron measurements. The single line observed at
200K arises then from an exchange or a motional mechanism which
narrows the g factor anisotropy. For the time scale of the ESR
measurements, at the highest frequency, the g anisotropy linewidth
can be evaluated at 1.7 T, which corresponds to a characteristic
time for the measurement of $20$ps. We conclude that the
electronic fluctuation time is temperature dependent with a value
smaller than 20ps above 100K while below 100K, it increases to a
value longer than 20ps. From $\mu$SR data in the paramagnetic
regime, the electronic fluctuation time is estimated at $32$ps for
\chem{LiNiO_2} at 100K; it is outside the muon time window for the
Mg doped sample at 50K. These results are quite consistent with
those of ESR, assuming that the Mg doped sample has a higher
fluctuating rate. In EXAFS the time scale of the measurement is
much shorter, typically $10^{-15}$s so that no narrowing process
is observed at room temperature. As for the neutron diffraction
measurements in ref\cite{CHU05}, we do not know the time scale of
the experiment so we cannot conclude further. We propose then the
following mechanism: the electrons occupy the $\vert
z^{2}-r^{2}\rangle$ orbitals but the \textit{z} axis moves. This
mechanism is reduced at low temperature at the same time when
magnetic excitations are developing (opening of a zero field gap).

These magnetic excitations are clearly observed in the Mg doped
sample as the progressive opening of a gap for both extremal
branches which follows the sample magnetization (measured at 1T).
The importance of the gap in branch 1 (51 GHz at 4K) rules out an
interpretation with a ferromagnetic origin. both branches look
rather like what is observed in an antiferromagnet. However no
long range magnetic order occurs in this sample as seen from the
magnetization and muon measurements. This means that the time
scale  on which the sample is probed in ESR is sufficiently short
to excite collective magnons. Looking at the lineshift due to
these magnons at 5K for instance in \chem{LiMgNiO_2}, 50GHz for
one branch, 5 GHz for the other branch, we get a characteristic
time at $2\times 10^{-11}$s and $2\times 10^{-10}$s respectively.
This means that the correlation time of the spin fluctuations is
longer than these values for magnons to be observed in ESR. In
$\mu$SR, if we assume the same dipolar field created by the Ni
ions as in \chem{NaNiO_2}, a precession at $10^{-7}$-$10^{-8}$s is
expected if the spin correlation time is longer. This is not the
case in our samples.

 If we now compare the ESR spectra for each sample,
we see that they have the same intrinsic line width, around 0.5 T,
as observed with $\mu$SR. The different magnon branches are better
defined in the Mg sample (the total lineshape takes the form of
two, quite well separated, modes)  than in \chem{LiNiO_2} (the
lineshape appears more like a continuous distribution of modes).
This may be related to the differences observed in the $\mu$SR
data: the quasistatic glassy state in \chem{LiNiO_2} gives rise to
a large distribution of magnetic modes; the different relaxing
phases in the Mg doped sample may be related to the two main
magnetic modes observed in ESR: one is associated with the $g_1$
value, and has the larger ESR signal. It corresponds to the 2/3
statistical weight of the $g_\perp$ contribution in a powder. It
gives the quasistatic signal in $\mu$SR. The second one is
associated with the $g_2$ value and corresponds to the 1/3
$g_\parallel$ contribution. It fluctuates two quickly to give a
contribution in $\mu$SR. At high temperature, the spins in the
parallel and perpendicular directions are coupled via an exchange
mechanism which narrows the ESR line and, at the same time, pushes
the spin fluctuation rate outside de $\mu$SR time window. In
LiNiO$_2$, this same exchange mechanism occurs at high
temperature, although probably at a slower rate since the spin
contribution is visible in $\mu$SR. At low temperature, the quasi
static state concerns the whole sample as seen in $\mu$SR.

\section{Conclusion}

In conclusion, we have studied a Mg doped \chem{LiNiO_2} sample
without interplane Ni ions compared it to a quasi stoichiometric
\chem{LiNiO_2} sample.
 Both the magnetization and $\mu$SR data clearly show that the samples do not undergo a
transition to long-ranged magnetic order while the ESR data
demonstrate the presence of magnetic excitations with a
correlation time longer than $10^{-11}$-$10^{-10}$s. This
low-temperature state does, however, change with Mg-doping. In our
quasi-stoichiometric \chem{LiNiO_2} sample a disordered, slowly
fluctuating state develops in the whole sample volume below
$12$~K. The corresponding antiferromagnetic magnons excited in ESR
have a large frequency distribution with a complex temperature
dependence. Mg-doping leads to faster electronic fluctuations and
smaller slowly fluctuating volume for $\mu$SR. The
antiferromagnetic magnons in ESR are better defined and the
largest spin gap follows the macroscopic magnetization.
 In the low temperate state, both compounds present an anisotropy of the g factor in
agreement with the $\vert z^{2}-r^{2}\rangle$ orbital occupancy
with $g_\parallel=2.0$and $g_\perp=2.4$. A motional narrowing
process occurs at the same time when the magnetic excitations
disappear, independent of the Mg doping. From this study, it is
clear that interplane Ni ions alter the magnetic properties of
\chem{LiNiO_2} but its removal and replacement by Mg is not
sufficient to allow the establishment of long range magnetic
order. In addition, we have shown that both compounds have a
single orbital occupancy, as in NaNiO$_2$, but an exchange
mechanism correlated with the magnetic interactions produces
dynamical effects above 100K.

\begin{acknowledgments}
Part of this work was performed at the Swiss Muon Source, Paul
Scherrer Institute, Villigen, Switzerland. We thank Alex Amato for
technical assistance. T.L. acknowledges support from the Royal
Commission  of 1851. We wish to acknowledge B\'alint N\'afr\'adi
for helping us with the ESR experiments and L\'aszl\'o Forr\'o
from the Swiss Federal Institute of Technology in
Lausanne/Switzerland for gratefully providing the ESR experimental
setup. The Grenoble High Magnetic Field Laboratory is associated
to the Universit\'{e} Joseph Fourier-Grenoble I.

\end{acknowledgments}

\bibliography{LiMgNiOv6}

\end{document}